\documentclass[12pt,a4paper]{revtex4}
\usepackage{amsmath}  
\usepackage{amssymb}  
\usepackage{dsfont}   

\begin{document}
\title{Coherent states for free particles}
\author{A. C. de la Torre}\email{delatorre@mdp.edu.ar}
\author{D. M. Goyeneche}\email{dgoyene@mdp.edu.ar}
 \affiliation{ IFIMAR - CONICET \\ Departamento de F{\'\i}sica -
 Facultad de Ciencias Exactas y Naturales \\
 Universidad Nacional de Mar del Plata\\
 Funes 3350, 7600 Mar del Plata, Argentina.}
\begin{abstract}
The coherent states are reviewed with particular application to
the free particle system. The didactic advantages of the
formalism is emphasized. Several interesting features, like the
relation of the coherent states with the Galilei group and with
the Husimi distribution are presented.
\end{abstract}
\maketitle
\section{INTRODUCTION}
Coherent states were discovered by Schr{\"o}dinger\cite{schr} in 1926
but remained unexploited until Glauber\cite{glau} rediscovered
them in 1963 and developed a profound formalism for them with
important applications; for instance, the ``Glauber states'' are
fundamental in the description of the electromagnetic field for
quantum optics. He was awarded the Nobel price in 2005 for the
development of the quantum theory of optical coherence. Besides
its important applications in quantum optics, theses states
provide an elegant mathematical description of the harmonic
oscillator and in many textbooks they are presented as a didactic
tool. However, the didactic advantage of the coherent states has
been ignored in the treatment of a free particle. It is, in fact,
very convenient to present the gaussian states of a free
particle, not as is usually done by assuming a gaussian state in
momentum space and (through Fourier transformation) obtaining the
corresponding gaussian in configuration space with a quite
unappealing mathematics, but instead, one can present the
coherent states for a free particle, and with simple and elegant
algebraic methods, derive all properties of the gaussian states.
The treatment of the coherent states of a free particle is
similar to the corresponding treatment of the harmonic oscillator
but there are remarkable differences and therefore it is
convenient to present the complete derivation that can be
directly used in the teaching of quantum mechanics in the
advanced undergraduate or graduate level.
\section{the free  PARTICLE system}
Let us consider the simplest quantum system consisting in one
free structureless particle moving in a one dimensional space.
Let $\mathcal{H}$ be the Hilbert space for the description of the
system. It is usual to choose the space of square integrable
functions $\mathcal{L}_{2}$ for this space but in principle it
can remain abstract. Position $X$ and momentum $P$ are the unique
independent observables of this system. Their eigenvectors
$\{\varphi_{x}\}$ and $\{\phi_{p}\}$, build two mutually unbiased
bases and their commutation relation is $[X,P]=i\hbar$. (We
assume here an infinite dimension of the Hilbert space; for
finite dimension, in a lattice for instance, the commutation
relation is more complicated\cite{dlt}). The hamiltonian of the
system is $H=\frac{P^{2}}{2m}$ and the time evolution operator is
given by $U_{t}=\exp(-\frac{i}{\hbar}Ht)$.
\section{ the coherent states}
It is very convenient to introduce a length scale $\lambda$ for
the system that will allow us, together with $\hbar$, to make
position and momentum dimensionless  in order to be able to add
them as simple numbers. For this scale we can choose whatever we
like, for instance the Compton length of the particle or any
measure of delocalization of the particle in space (we will later
see that it is convenient to choose $\lambda$ proportional to the
width of the space distribution). With this scale we define the
operator $A$ and its hermitian adjoint
\begin{eqnarray}
  A &=& \frac{1}{\lambda\sqrt{2}}\ X +
  i\frac{\lambda }{\hbar\sqrt{2}}\ P \ ,\\
  A^{\dag} &=& \frac{1}{\lambda\sqrt{2}}\ X -
  i\frac{\lambda }{\hbar\sqrt{2}}\ P \ .
\end{eqnarray}
These operators are dimensionless and their commutation relation
is
\begin{equation}\label{com}
  [A,A^{\dag}]=\mathbb{I}\ .
\end{equation}
The relations above can be trivially inverted to obtain
\begin{eqnarray}
  X &=& \frac{\lambda }{\sqrt{2}}\left(A+ A^{\dag} \right)\ , \\
 P &=& \frac{\hbar}{i\sqrt{2}\lambda }\left(A- A^{\dag} \right) \ .
\end{eqnarray}
 Now we can
define the \emph{coherent states} $\psi_{\alpha}$ as the
eigenvectors of $A$:
\begin{equation}\label{coh}
 A\psi_{\alpha}=\alpha\psi_{\alpha}\ .
\end{equation}
Since $A$ is not hermitian its eigenvalues are complex:
$\alpha\in\mathbb{C}$. It is remarkable that we can very easily
calculate these eigenvalues, related with observable properties
of an ensemble of particles in this coherent state. Let us find
$\langle X \rangle,\langle P \rangle,\langle X^{2}
\rangle,\langle P^{2} \rangle$ and the widths $\Delta_{x}$ and
$\Delta_{p}$ for the particle in a coherent state
$\psi_{\alpha}$.
\begin{equation}\label{xm}
  \langle X\rangle =\frac{\lambda }{\sqrt{2}}
  \left(\langle\psi_{\alpha},A\psi_{\alpha}\rangle+
  \langle\psi_{\alpha},A^{\dag}\psi_{\alpha}\rangle \right)
  =\frac{\lambda }{\sqrt{2}}
  \left(\langle\psi_{\alpha},A\psi_{\alpha}\rangle+
  \langle A\psi_{\alpha},\psi_{\alpha}\rangle \right) =
  \frac{\lambda }{\sqrt{2}}
  \left(\alpha+  \alpha^{\ast} \right)\ ,
\end{equation}
and similarly we obtain
\begin{equation}\label{pm}
  \langle P\rangle =\frac{\hbar}{i\sqrt{2}\lambda }
  \left(\alpha- \alpha^{\ast} \right) \ .
\end{equation}
Adding them we obtain the eigenvalue $\alpha$ given in terms of
the expectation values of position and momentum
\begin{equation}\label{alf}
  \alpha=\frac{1}{\sqrt{2}\lambda}\langle X\rangle +
  i \frac{\lambda}{\sqrt{2}\hbar }\langle P\rangle  \ .
\end{equation}
We could have obtained this relation simply by taking the
expectation values in Eq.(1). The coherent states are then
completely determined by the expectation values of position and
momentum. For $\langle X^{2}\rangle $ we have
\begin{eqnarray}
\nonumber
 \langle X^{2}\rangle &=& \frac{\lambda^{2 }}{2}\langle\left(A+ A^{\dag}
 \right)^{2}\rangle=\frac{\lambda^{2 }}{2}\left\langle A^{2}+A^{\dag 2}
 +AA^{\dag}+A^{\dag}A\right\rangle \\&=&
 \frac{\lambda^{2 }}{2}\left\langle A^{2}+A^{\dag 2}
 +2A^{\dag}A+1\right\rangle=
  \frac{\lambda^{2 }}{2}\left( \alpha^{2}+\alpha^{\ast 2}
 +2\alpha^{\ast}\alpha+1\right)\ ,
\end{eqnarray}
and similarly we get
\begin{equation}\label{p2}
 \langle P^{2}\rangle =
  -\frac{\hbar^{2}}{2\lambda^{2} }\left( \alpha^{2}+\alpha^{\ast 2}
 -2\alpha^{\ast}\alpha+1\right)\ .
\end{equation}
The widths of the position and momentum distributions are then
\begin{eqnarray}
  \Delta^{2}_{x}&=&\langle X^{2}\rangle-\langle X\rangle^{2}=
\frac{\lambda^{2}}{2}\ , \\
\Delta^{2}_{p}&=&\langle P^{2}\rangle-\langle P\rangle^{2}=
\frac{\hbar^{2}}{2\lambda^{2} }\ ,
\end{eqnarray}
where  we see that the length scale is related to the width in
position and  we get the important result that, in a coherent
state, the uncertainty relation is optimized
$\Delta_{x}\Delta_{p}=\frac{\hbar}{2}$. For this reason, it is
often mentioned that the coherent states are the quantum states
closest to classical behaviour. In these coherent states,
position and momentum have the least correlation (the real part
of the quantum covariance function\cite{dlt3} vanishes) in the
sense that $\langle XP+PX\rangle-2\langle X\rangle\langle
P\rangle =0$.

We have seen that the eigenvectors $\psi_{\alpha}$ of the
operator $A$ have very interesting properties and we may wonder
whether the eigenvectors of $A^{\dag}$ may be of some interest.
We can see however that the eigenvalue equation
$A^{\dag}\chi_{\beta}=\beta\chi_{\beta}$ does not have solutions
because, if $\chi_{\beta}$ would exist then we would obtain an
absurd result. In fact, if we assume that $\chi_{\beta}$ exists,
applying the same techniques as before, we can calculate $\langle
X \rangle, \langle X^{2} \rangle$ and the width
$\Delta^{2}_{x}=\langle X^{2} \rangle-\langle X \rangle^{2}$ for
the particle in the state $\chi_{\beta}$. Doing this we obtain
$\Delta^{2}_{x}=-\lambda^{2}/2$; a negative result! This is of
course impossible because $\Delta^{2}_{x}$ is the expectation
value of the \emph{positive} operator $(X-\langle X
\rangle)^{2}$, therefore there is no $\chi_{\beta}$ satisfying
$A^{\dag}\chi_{\beta}=\beta\chi_{\beta}$.
\section{time evolution of coherent states}
So far, the treatment of the coherent states for a free particle
and for the harmonic oscillator are similar and we can now
present some important differences. In the harmonic oscillator,
one defines a ``number operator'' $N=AA^{\dag}$ having
nonnegative integer eigenvalues, with $A$ and $A^{\dag}$ as shift
operators for the eigenvectors of $N$. This operator is very
important for the harmonic oscillator because it has a relevant
physical meaning; it is, essentially, the hamiltonian operator
that controls the time evolution. Of course we can also define
this number operator for the free particle, and we will do so in
the next section, but in this case it is not related to a
physically interesting observable. In the free particle case, the
hamiltonian is given by the momentum squared and the energy
eigenvectors are the same as the momentum eigenvectors $\phi_{p}$
(notice that the energy eigenvalue $E=p^{2}/2m$ has  a twofold
degeneracy with eigenvectors $\phi_{p}$ and $\phi_{-p}$.)

If at time $t=0$ the free particle is in the coherent state
$\psi_{\alpha}$ then at time $t$ it will be in the state
$\Psi(t)$ given by the time evolution operator
\begin{equation}\label{tevol}
   \Psi(t)=\exp\left(-\frac{i}{\hbar}Ht\right)\ \psi_{\alpha} =
   \exp\left( \frac{ -it}{ 2m\hbar}P^{2}\right)\ \psi_{\alpha}=
   \exp\left( \frac{  i\hbar t}{ 4m\lambda^{2}} (A- A^{\dag})^{2}
   \right)\ \psi_{\alpha}\ .
\end{equation}
Here we can see another important difference with the harmonic
oscillator. One can prove that the coherent states of the
harmonic oscillator remain coherent through the time evolution
whereas in the free particle case the time evolution destroys the
coherent states. That is, $\Psi(t)$, for $t\neq 0$, is no longer
a coherent state. This is physically understood because, as is
well known,
 the width in the position distribution, $\Delta_{x}$,
increases with time for the free particle and the width in
momentum, $\Delta_{p}$, remains constant (because the momentum
distribution is time independent); therefore the uncertainty
product $\Delta_{x}\Delta_{p}$ will also increase with time and
will no longer have the minimal value $\hbar/2$ characteristic of
the coherent states. Anyway we will now prove that $\Psi(t)$ is
\emph{not} an eigenvector of $A$ and therefore it is not a
coherent state. For this, consider the time evolution applied to
the Eq.(\ref{coh}) that results in
\begin{equation}\label{tevol}
\exp\left(-\frac{i}{\hbar}Ht\right)\ A\
\exp\left(\frac{i}{\hbar}Ht\right)\Psi(t) = \alpha\Psi(t)\ .
\end{equation}
Now, for the operator in the left hand side of this equation we
can use the general expression valid for any constant $\gamma$
and any operators $A$ and $B$
\begin{equation}\label{tevo11}
 \exp\left(-\gamma B\right)\ A\
\exp\left(\gamma B\right)= A + \gamma [B,A] +
\frac{\gamma^{2}}{2!}[B,[B,A]] + \cdots
+\frac{\gamma^{n}}{n!}[B,[B,\cdots ,A]]\cdots ] + \cdots
\end{equation}
To prove this equation, consider the function  $f(\gamma)=
\exp\left(-\gamma B\right)A\exp\left(\gamma B\right)$ and
calculate the derivatives with respect to $\gamma$. Now making a
Taylor series expansion about $\gamma =0$ we obtain the relation
above. In our case, however, the series is interrupted after the
second term:
\begin{equation}\label{tevol2}
 [H,A]=\frac{1}{2m}[P^{2},A]=\frac{1}{2\sqrt{2}m\lambda}[P^{2},X]
 =\frac{-i\hbar}{\sqrt{2}m\lambda}P\ ,
\end{equation}
and $[H,[H,A]]=0$. Therefore Eq.(\ref{tevol}) becomes
\begin{equation}\label{tevol3}
\left(A+\frac{t}{\sqrt{2}m\lambda}P\right)\Psi(t) =
\alpha\Psi(t)\ ,
\end{equation}
that is,
\begin{equation}\label{tevol4}
A\Psi(t)=\frac{-t}{\sqrt{2}m\lambda}P\Psi(t) + \alpha\Psi(t)\neq
\beta\Psi(t)\ .
\end{equation}
The last inequality follows because $\Psi(t)$ is \emph{not} an
eigenvector of $P$ (neither is $\psi_{\alpha}$) and therefore it
is also not an eigenvector of $A$.
\section{The number operator}
As was announced at the beginning of last section, we will now
study the number operator $N=AA^{\dag}$ that has interesting
formal properties although it is not related to any important
observable of the free particle physical system, whose most
relevant observables are position and momentum. Another operator
similar to $N$ is $A^{\dag}A$ and one may wonder whether it may
bring something interesting. However, due to the commutation
relation $[A,A^{\dag}]=\mathbb{I}$, it is clear that this
operator is not significantly different from $N$ and it amounts
only to the addition of a constant. The operator $N$ is hermitian
and positive $N\geq 1$, that is, it is bounded from bellow. This
is important because one can prove that for such an operator
there exist eigenvalues and eigenvectors
$(\lambda_{n},\chi_{n})$. The prove that $N^{\dag}=N$ is trivial
and the positivity follows from
\begin{equation}\label{Npos}
\langle\Psi,AA^{\dag}\Psi\rangle = \langle
A^{\dag}\Psi,A^{\dag}\Psi\rangle = \|A^{\dag}\Psi\|^{2} \geq 0\ ,
\forall \Psi\in \mathcal{H}\ .
\end{equation}
Then we have $N\chi_{n}=\lambda_{n}\chi_{n}$, with
$\lambda_{n}\geq 0$. Using the commutation relations
\begin{equation}\label{commNA}
[N,A^{\dag}]=A^{\dag} \mbox{ and }[N,A]=-A
\end{equation}
one can easily prove that $A^{\dag}\chi_{n}$ is another
eigenvector of $N$ but corresponding to the eigenvalue
$\lambda_{n}+1$ and $A\chi_{n}$ is another eigenvector of $N$ but
corresponding to the eigenvalue $\lambda_{n}-1$. Now, if we start
with an arbitrary eigenvector $\chi_{n}$ and apply $A$ a
sufficient large number of times, then we would get an
eigenvector with a \emph{negative} eigenvalue in contradiction
with the positivity of $N$. The solution of this difficulty is
that the eigenvalues must be integer numbers. Then we can set
$\lambda_{n}=n = 0, 1, 2, \ldots $, and there is a \emph{lowest
stair} $\chi_{0}$ such that $A\chi_{0}=0$ from which we can
obtain any $\chi_{n}$ by applying $n$ times the raising operator
$A^{\dag}$. Summarizing, we have:
\begin{eqnarray}
N\chi_{n}&=&  n \chi_{n} \mbox{ with } n = 0, 1, 2, \ldots \\
A^{\dag}\chi_{n}&=&  \sqrt{n+1}\ \chi_{n+1} \ ,\\
A \chi_{n}&=&  \sqrt{n }\ \chi_{n-1} \mbox{ , in particular, } A \chi_{0}=0\ ,\\
\chi_{n} &=& \frac{1}{\sqrt{n!}}(A^{\dag})^{n}\chi_{0}\ .
\end{eqnarray}
Notice that from Eqs.(\ref{coh} and 24) it follows that the
lowest number state coincides with the coherent state for $\alpha
=0$, that is, $\chi_{0}=\psi_{0}$. We will later see how every
coherent state $\psi_{\alpha}$ can be expanded in the number
states basis $\{\chi_{n}\}$.
\section{The drift operator}
An interesting operator for the coherent states formalism is the
drift operator defined as
\begin{eqnarray}
\nonumber D(\alpha) &=&\exp\left(\alpha A^{\dag}-\alpha^{\ast}
A\right)
\\ \nonumber &=&
 \exp\left(-\frac{|\alpha |^{2}}{2}\right)
 \exp\left(\alpha A^{\dag}\right)\exp\left(-\alpha^{\ast}
 A\right)\\
&=& \exp\left(\frac{|\alpha |^{2}}{2}\right)
\exp\left(-\alpha^{\ast} A\right) \exp\left(\alpha
A^{\dag}\right)\ .
\end{eqnarray}
The last  two expressions for $D(\alpha)$ follow from the
Hausdorff-Baker-Campbell (HBC) relation:
\begin{equation}\label{hbc}
\exp(R)\exp(S)=\exp\left(R+S\right)\exp\left(\frac{[R,S]}{2}\right)
\ ,\mbox{  valid if  } [R,[R,S]]=[S,[R,S]]=0
   \ .
\end{equation}
Repeated application of this identity results in
\begin{equation}\label{hbc1}
\exp(R)\exp(S)=\exp(S)\exp(R)\exp([R,S]) \ ,\mbox{ valid if  }
[R,[R,S]]=[S,[R,S]]=0
   \ .
\end{equation}
With the HBC relation, one can also prove that the drift operator
is unitary,
$D^{\dag}(\alpha)D(\alpha)=D(\alpha)D^{\dag}(\alpha)=\mathbb{I}$,
therefore $D^{\dag}(\alpha)=D^{-1}(\alpha)=D(-\alpha)$. The set
$\{D(\alpha), \forall \alpha\in\mathbb{C}\}$ is a group whose
product can also be calculated with the HBC relation as
\begin{equation}\label{grupo}
D(\alpha)D(\beta)=
\exp\left(\frac{\alpha\beta^{\ast}-\alpha^{\ast}\beta}{2}\right)\
D(\alpha+\beta)\ .
\end{equation}
 We will prove that the
operator $D(\alpha)$ can be used to relate coherent states
corresponding to different values of $\alpha$. More precisely, if
$\psi_{0}$ is the coherent state for $\alpha=0$ then we have
\begin{equation}\label{drift1}
\psi_{\alpha}= D(\alpha)\psi_{0}\ .
\end{equation}
In order to  prove this, we will need some algebraic properties
of the operators $A$ and $A^{\dag}$. By mathematical induction
one can prove that, from the commutation relation
$[A,A^{\dag}]=\mathbb{I}$, it follows that $[A, A^{\dag n}]= n
A^{\dag (n-1)}$, and from this, we can show that for  every
function $F$ that can be expanded as a power series it is
\begin{equation}\label{drift2}
[A, F(A^{\dag})]=  \frac{dF(A^{\dag})}{dA^{\dag}}\ .
\end{equation}
Applying this to the drift operator we have
\begin{equation}\label{drift3}
[A, D(\alpha)]=  \alpha D(\alpha)\ .
\end{equation}
With this result we can easily see that $D(\alpha)\psi_{0}$ is an
eigenvector of $A$ corresponding to the eigenvalue $\alpha$. In
fact, we have:
\begin{equation}\label{drift4}
AD(\alpha)\psi_{0}=D(\alpha)A\psi_{0} + \alpha D(\alpha)\psi_{0}
=\alpha D(\alpha)\psi_{0}\ .
\end{equation}
Therefore $D(\alpha)\psi_{0}= k\psi_{\alpha}$, where $k$ is some
proportionality constant. In order to determine it, consider the
norm of the equation
\begin{equation}\label{drift5}
|k|^{2}\langle \psi_{\alpha},\psi_{\alpha}\rangle=\langle
D(\alpha)\psi_{0},D(\alpha)\psi_{0}\rangle
=\langle\psi_{0},D^{\dag}(\alpha)D(\alpha)\psi_{0}\rangle
=\langle\psi_{0},\psi_{0}\rangle = \|\psi_{0}\|^{2}\ .
\end{equation}
Assuming that the coherent states are normalized, $\langle
\psi_{\alpha},\psi_{\alpha}\rangle=\|\psi_{\alpha}\|^{2}=1$, then
$|k|^{2}=1$ and $D(\alpha)\psi_{0}$ and $\psi_{\alpha}$ can
differ by a constant phase that can be set equal to $1$ and
therefore we obtain Eq.(\ref{drift1}). Notice that the algebraic
proof given here is different from the usual proof that involves
the expansion of the coherent states in the basis of the
eigenvectors of the number operator. In this approach we don't
use such a basis because they are physically uninteresting for
the free particle system and we prefer to rely only in the
algebraic structure of the operators. The drift operator has an
intuitive meaning that becomes evident when we write $D(\alpha)$
in terms of the position and momentum operators: Using Eqs.(1, 2,
9, 26), we have
\begin{eqnarray}
\nonumber D(\alpha) &=&\exp\left(\frac{i}{\hbar}(\langle P\rangle
X-\langle X\rangle P)\right) \\ \nonumber &=&
 \exp\left(-\frac{i}{2\hbar}\langle
X\rangle \langle P\rangle\right)
 \exp\left( \frac{i}{\hbar} \langle P\rangle X\right)
 \exp\left(-\frac{i}{\hbar} \langle X\rangle P \right) \\
&=&\exp\left( \frac{i}{2\hbar}\langle X\rangle \langle
P\rangle\right) \exp\left(- \frac{i}{\hbar} \langle X\rangle
P\right) \exp\left(\frac{i}{\hbar} \langle P\rangle X\right)\ ,
\end{eqnarray}
and recalling that $\exp\left(-\frac{i}{\hbar} a P\right)$ is a
translation operator that performs $X\rightarrow X+a$ and
$\exp\left( \frac{i}{\hbar} g X\right)$ changes the momentum
$P\rightarrow P+g$, we see that $D(\alpha)$ transforms a coherent
state corresponding to a particle at rest, $\langle P\rangle=0$,
located at the origin, $\langle X\rangle=0$, into a coherent
state for the particle at position $\langle X\rangle$ and moving
with momentum $\langle P\rangle$. Notice that the different
choice of what transformation is performed first, amounts only to
a different constant phase. It is then interesting to notice
that, effectively, $D(\alpha)$ performs a Galilei transformation
on the state $\psi_{0}$ and therefore the group build with the
drift operators $D(\alpha)$ is isomorph with the Galilei group.
In fact, the phase appearing in Eq.(\ref{grupo}) shows that this
group is a \emph{projective} group (not simple) as is the case
with the Galilei group. The emergence of the Galilei group is an
additional ``bonus'' of the formalism of coherent states.

The arguments presented in this section are valid for a free
particle as well as for the harmonic oscillator, or for any other
potential. The only difference is that, in the harmonic
oscillator case, the state $\psi_{0}$ has the unique feature of
being the ground state of the Hamiltonian and one may prefer to
choose this state in order to generate all other coherent states.
In the free particle case, $\psi_{0}$ is a state as good as any
other because of the Galilei invariance of the free particle
system and one might prefer to write Eq.(\ref{drift1}) in an
unbiased way as $\psi_{\beta}=D(\beta-\alpha)\psi_{\alpha}$.
\section{ coherent states are gaussian}
One of the most beautiful features of the Hilbert space formalism
of quantum mechanics is that almost all results can be obtained
without specifying  a particular representation of the Hilbert
space. Everything follows from the algebraic and geometric
structure of the Hilbert space. Some features become however
easier to grasp in some particular representation of the Hilbert
space. For instance if we are interested in the position
distribution of a particle it becomes natural to choose the space
$\mathcal{L}_{2}$ of square integrable functions of a real
variable $x$. As is well known, the position operator $X$ in this
space is the multiplication by $x$ and the momentum operator $P$
is the derivative $-i\hbar\frac{d}{dx}$. In order to find the
coherent states
$\psi_{\alpha}(x)=\langle\varphi_{x},\psi_{\alpha}\rangle$ in
this space we must write and solve Eq.(\ref{coh}) in
$\mathcal{L}_{2}$. That is
\begin{equation}\label{cohl2}
  \left[\frac{1}{\lambda\sqrt{2}}\ x +
  i\frac{\lambda }{\hbar\sqrt{2}}\ (-i\hbar\frac{d}{dx})  \right]
  \psi_{\alpha}(x)=\alpha\psi_{\alpha}(x)\ .
\end{equation}
In terms of a dimensionless variable $y=x/(\sqrt{2}\lambda)$ this
becomes
\begin{equation}\label{cohl21}
 \frac{d\psi_{\alpha}(y)}{dy} = 2(\alpha-y)\psi_{\alpha}(y)\ ,
\end{equation}
that is,
\begin{equation}\label{cohl22}
 \frac{d\psi_{\alpha}}{\psi_{\alpha}} = 2(\alpha-y)dy\ .
\end{equation}
Integrating we have
\begin{equation}\label{cohl23}
 \ln(\psi_{\alpha}) = -y^{2}+2\alpha y +\mathrm{Const.}\ ,
\end{equation}
that is,
\begin{equation}\label{cohl23}
\psi_{\alpha}(y) = C\exp\left(-y^{2} +2\alpha y\right) \ .
\end{equation}
The next step is to complete the square in the exponential and
absorb the corresponding factor in the constant $C$ (that will be
anyway determined by normalization). After replacing the original
variable $x$ and $\alpha$ given by Eq.(\ref{alf}) with $\langle
X\rangle=x_{0}$ and $\langle P\rangle=p_{0}$, and replacing the
length scale $\lambda=\sqrt{2}\Delta_{x}$ we obtain
\begin{equation}\label{cohl24}
\psi_{\alpha}(x) =
\left[2\pi\Delta_{x}^{2}\right]^{-1/4}\exp\left(
-\frac{(x-x_{0})^{2}}{4\Delta_{x}^{2}}+i
\frac{p_{0}}{\hbar}x\right) \ .
\end{equation}
We see therefore that the coherent states of a free particle are
the gaussian states (in order to appreciate the didactic
advantages of the coherent states for a free particle, compare
this derivation with the presentation of the of gaussian wave
packets as is usually done in textbooks).

One last remark in this section is that the coherent states in
momentum representation, that is, the functions
$\psi_{\alpha}(p)=\langle\phi_{p},\psi_{\alpha}\rangle$ are also
gaussian. There is in fact a symmetry in the exchange of position
and momentum  for the coherent states.
\section{Expansion in number states and in coherent states}
A very important feature of the Hilbert space formalism is the
fact that the eigenvectors of an hermitian operator, that may be
related to a physical observable, build a basis. If we expand the
state $\Psi$ of a system in this basis, we may interpret the
modulus squared of the expansion coefficients as the
probabilities associated with the eigenvalues of the observable.

A relevant question is then, if we can use the set of coherent
states $\{\psi_{\alpha}\}$ in order to make an expansion of a
state $\Psi$ and interpret the expansion coefficients
$|\langle\psi_{\alpha},\Psi\rangle|^{2}$ as some probability
distribution. The first remark in order to answer this, is that
the coherent states are the eigenvectors of an operator that is
\emph{not} hermitian and they do not build an orthonormal basis.
They are not even linearly independent (if they were independent
we could transform them into a basis by Schmidt orthogonalization
procedure). We will show that the coherent states are not
orthogonal, that is,
$\langle\psi_{\alpha},\psi_{\beta}\rangle\neq 0$.
\begin{equation}\label{prod}
\langle\psi_{\alpha},\psi_{\beta}\rangle =\langle
D(\alpha)\psi_{0},D(\beta)\psi_{0}\rangle \ .
\end{equation}
Using Eq.(26) and considering that $\exp(aA)\psi_{0}=\psi_{0}$ we
get
\begin{eqnarray}
\nonumber \left\langle\psi_{\alpha},\psi_{\beta}\right\rangle &=&
\left\langle \exp\left(-\frac{|\alpha|^{2}}{2}\right)
\exp\left(\alpha
A^{\dag}\right)\psi_{0},\exp\left(-\frac{|\beta|^{2}}{2}\right)
\exp\left(\beta A^{\dag}\right)\psi_{0}\right\rangle \\
  &=&
\exp\left(-\frac{|\alpha|^{2}}{2}-\frac{|\beta|^{2}}{2}\right)
\left\langle\psi_{0},\exp\left(\alpha^{\ast}
A\right)\exp\left(\beta A^{\dag}\right)\psi_{0}\right\rangle\ .
\end{eqnarray}
Now, using Eq.(\ref{hbc1}) we can permute the operators and we
get
\begin{eqnarray*}
\nonumber \left\langle\psi_{\alpha},\psi_{\beta}\right\rangle &=&
\exp\left(-\frac{|\alpha|^{2}}{2}-\frac{|\beta|^{2}}{2}+\alpha^{\ast}\beta\right)
\left\langle\psi_{0},\exp\left(\beta
A^{\dag}\right)\exp\left(\alpha^{\ast}
A\right)\psi_{0}\right\rangle\\
&=&
\exp\left(-\frac{|\alpha|^{2}}{2}-\frac{|\beta|^{2}}{2}+\alpha^{\ast}\beta\right)
\left\langle\psi_{0},\psi_{0}\right\rangle\\
&=&\exp\left(-\frac{|\alpha|^{2}}{2}-\frac{|\beta|^{2}}{2}+
\alpha^{\ast}\beta\right)\ .
\end{eqnarray*}

The three operators $X$, $P$, $N$ define tree bases of the
Hilbert space $\{\varphi_{x}\}$, $\{\phi_{p}\}$ and
$\{\chi_{n}\}$. The first two are continuous $x, p \in
\mathbb{R}$ and unbiased, that is,
 $|\langle\varphi_{x},\phi_{p}\rangle|$ is a constant independent
of $x$ and $p$. The third basis is numerable and biased with the
other two. One can calculate, for instance, that
$\langle\varphi_{x},\chi_{n}\rangle = \chi_{n}(x)$ is related
with the Hermite polynomials. We will not reproduce here this
calculation that can be found in many quantum mechanics
textbooks. Instead of this, we are interested in the expansion of
the coherent states in these bases. We have already seen in
section VII that the Fourier coefficients of the expansion of the
coherent states in the bases of position and momentum, that is
$\langle\varphi_{x},\psi_{\alpha}\rangle$ and
$\langle\phi_{p},\psi_{\alpha}\rangle$, are gaussian functions.
The corresponding coefficients,
$\langle\chi_{n},\psi_{\alpha}\rangle$, of the expansion of the
coherent states in this new basis $\{\chi_{n}\}$ are given by
\begin{multline}\label{coeff}
 \langle\chi_{n},\psi_{\alpha}\rangle  =
\langle\frac{1}{\sqrt{n!}}(A^{\dag})^{n}\chi_{0},\psi_{\alpha}\rangle
= \frac{1}{\sqrt{n!}}\langle \chi_{0},A^{n}\psi_{\alpha}\rangle =
\\
\frac{1}{\sqrt{n!}}\langle
\psi_{0},\alpha^{n}\psi_{\alpha}\rangle =
\frac{\alpha^{n}}{\sqrt{n!}}\langle \psi_{0},\psi_{\alpha}\rangle
=
\frac{\alpha^{n}}{\sqrt{n!}}\exp\left(-\frac{|\alpha|^{2}}{2}\right)\
.
\end{multline}
We have  therefore
\begin{equation}\label{expan}
\psi_{\alpha} = \sum_{n}
\frac{\alpha^{n}}{\sqrt{n!}}\exp\left(-\frac{|\alpha|^{2}}{2}\right)\
\chi_{n}\ .
\end{equation}
We can now go back to our question of whether it is possible to
make an expansion of any state in terms of the coherent states.
We have seen that these states are not orthogonal; however they
build an (over)complete set of states in the sense that
\begin{equation}\label{overcomp}
\pi^{-1}\int_{\alpha\in\mathbb{C}}\!\!\!\!\!\!d^{2}\alpha\
\psi_{\alpha} \langle\psi_{\alpha},\cdot\rangle = \mathbb{I}\
\end{equation}
(we use here the correct Hilbert space notation for a projector:
 $\Psi\langle\Psi,\cdot\rangle$; for those addict to
the Dirac notation this is $|\Psi\rangle\langle\Psi|$) and we can
use this relation in order to expand any element $\Psi$ as
\begin{equation}\label{expand}
\Psi=\mathbb{I}\Psi=\pi^{-1}\int_{\alpha\in\mathbb{C}}\!\!\!\!\!\!d^{2}\alpha\
\psi_{\alpha} \langle\psi_{\alpha},\Psi\rangle\ .
\end{equation}
The proof of the completeness relation in Eq.(\ref{overcomp})
follows from the completeness of the basis $\{\chi_{n}\}$:
\begin{eqnarray}
\nonumber \int_{\alpha\in\mathbb{C}}\!\!\!\!\!\!d^{2}\alpha\
\psi_{\alpha} \langle\psi_{\alpha},\cdot\rangle &=&
\int_{\alpha\in\mathbb{C}}\!\!\!\!\!\!d^{2}\alpha\
\sum_{n}\frac{\alpha^{n}}{\sqrt{n!}}\exp\left(-\frac{|\alpha^{2}|}{2}\right)
\sum_{m}\frac{\alpha^{\ast
m}}{\sqrt{m!}}\exp\left(-\frac{|\alpha^{2}|}{2}\right)
\chi_{n}\langle\chi_{m},\cdot\rangle \\ \nonumber
  &=& \sum_{n,m}
\chi_{n}\langle\chi_{m},\cdot\rangle \frac{1}{\sqrt{n!m!}}
\int_{\alpha\in\mathbb{C}}\!\!\!\!\!\!d^{2}\alpha\ \alpha^{n}
\alpha^{\ast m} \exp\left(-|\alpha^{2}|\right) \\ \nonumber
  &=&  \sum_{n,m}
\chi_{n}\langle\chi_{m},\cdot\rangle \frac{1}{\sqrt{n!m!}} \pi n!\delta_{n,m} \\
  &=&\pi\sum_{n}\chi_{n}\langle\chi_{n},\cdot\rangle  = \pi \mathbb{I}\ .
\end{eqnarray}
The integral in the complex plane is performed in polar
representation $\alpha = r \exp(i\theta)$.

The completeness relation in Eq.(\ref{overcomp}) can be used to
expand any Hilbert space element. For instance, we can invert
Eq.(\ref{expan}) and give the number states expanded in the
coherent states, using Eq.(\ref{coeff}):
\begin{equation}\label{expandNumbSt}
\chi_{n}=\mathbb{I}\chi_{n}=\pi^{-1}\int_{\alpha\in\mathbb{C}}
\!\!\!\!\!\!d^{2}\alpha\
\frac{\alpha^{*n}}{\sqrt{n!}}\exp\left(-\frac{|\alpha|^{2}}{2}\right)
\psi_{\alpha} \ .
\end{equation}
This expression, when written in the Hilbert space of the squared
integrable functions, $\mathcal{L}_{2}$, allows an interesting
representation of the Hermite polynomials in terms of gaussian
functions.

One last comment concerning the expansion in terms of the
coherent states is that the set of elements $\{\psi_{\alpha}\}$,
as well as the basis elements $\{\chi_{n}\}$, are well defined in
the Hilbert space. Strictly speaking, the bases associated with
position and momentum $\{\varphi_{x}\}$ and $\{\phi_{p}\}$ do not
belong to the Hilbert space. In fact, if we choose the Hilbert
space of the square integrable functions of some real variable,
$y$, in the position representation where $ \varphi_{x}(y)=\delta
(x-y)$ and $\phi_{p}(y)=
\frac{1}{\sqrt{2}\hbar}\exp(\frac{i}{\hbar}py)$, we can
immediately see that they do not belong to the Hilbert space
because they are not squared integrable. For this reason, in
advanced quantum mechanics books, the Hilbert space is extended
to the \emph{rigged Hilbert space} that includes these elements
that may be used as bases. This is precisely what we do when we
calculate a Fourier transformation. The mathematics to justify
the extension of the Hilbert space to the rigged Hilbert space is
well explained in Chapter 1 of reference \cite{ball}.

\section{the Husimi distribution}
The main interest  in the expansion of a state $\Psi$ of a system
in terms of the eigenvector of some observables is that the
modulus squared of the expansion coefficient have the physical
interpretation as the probability distribution of the
corresponding eigenvalues. So,
$\rho(x)=|\langle\varphi_{x},\Psi\rangle|^{2}$ and
$\varpi(p)=|\langle\phi_{p},\Psi\rangle|^{2}$ are understood as
the probability distributions for position and momentum (although
this is perhaps a missnommer\cite{dlt1}). Similarly, we may try
to interpret $|\langle\psi_{\alpha},\Psi\rangle|^{2}$ as some
probability distribution for $\alpha$. From the mathematical
point of view, this seems to be possible because, using the
over-completeness relation Eq.(\ref{overcomp}) one can prove that
the expectation value of any function $F(A)$ can be given as
\begin{equation}\label{expvalF}
 \langle\Psi,F(A)\Psi\rangle = \langle\Psi,F(A)\mathbb{I}\Psi\rangle =
 \pi^{-1}\int_{\alpha\in\mathbb{C}}\!\!\!\!\!\!d^{2}\alpha\ F(\alpha)\
 |\langle\psi_{\alpha},\Psi\rangle|^{2}\ ,
\end{equation}
and therefore $\pi^{-1}|\langle\psi_{\alpha},\Psi\rangle|^{2}$
can be interpreted as the probability for the realization of the
value $\alpha$. However, from the physical point of view this
does not seem to make sense. The reason for this is that $A$ is
not an hermitian operator and therefore the complex eigenvalues
$\alpha$ can not be associated with the result of a single
measurement of some observable. Consequently, it is meaningless
to say that the system \emph{has} some value of $\alpha$ with
some probability. In other words, let us consider the complex
variable $\alpha$ as a function of two real variables $x$ and $p$
 as in Eq.(\ref{alf})
$\alpha(x,p)=\frac{1}{\sqrt{2}\lambda}x + i
\frac{\lambda}{\sqrt{2}\hbar }p $. Clearly, $x$ and $p$ can not
be the result of any measurement of position and momentum because
these can not be measured simultaneously. Neither is there an
observable, with its corresponding hermitian operator, whose
single measurement provides simultaneously the expectation values
of position and momentum. These expectation values can only be
obtained in a multiplicity of measurements in an ensemble of
identically prepared systems but not in some single measurement
of one system. More precisely, we can see that given an
\emph{arbitrary} state $\Psi$, that is, \emph{any} Hilbert space
element, with position expectation value $\langle
X\rangle=\langle\Psi,X\Psi\rangle$,  then there exists no
hermitian operator $\tilde{X}$ such that $\tilde{X}\Psi=\langle
X\rangle\Psi$. Clearly this is impossible because if such an
operator would exist, then \emph{all} Hilbert space elements
would belong to the basis associated with $\tilde{X}$.

Even though $|\langle\psi_{\alpha},\Psi\rangle|^{2}$ is not a
probability density, this quantity has been studied in one of the
several attempts to establish a compound probability distribution
for position and momentum (see for instance Chapter 15 of
reference \cite{ball}). That is, some function of $x$ and $p$
such that when integrated over $p$ results in the position
distribution $\rho(x)$ and when integrated over $x$ delivers the
momentum distribution $\varpi(p)$. The most famous attempt is the
Wigner distribution that has these two properties but is not
everywhere nonnegative. The function $\rho_{H}(x,p)=
\pi^{-1}|\langle\psi_{\alpha(x,p)},\Psi\rangle|^{2}$ is known as
the Husimi distribution that is everywhere nonnegative but, when
integrated over one of its variables, does not provide the
correct distribution for the other one. There are interesting
properties for this phase space ``distribution'' that can be
found in ref.\cite{ball} and a profound comparative study and
applications of many phase space distribution functions is found
in ref.\cite{lee}.

\section{CONCLUSIONS}
We have seen that the coherent states for a free particle, very
well known for the harmonic oscillator case and in quantum
optics, provide a very useful didactic tool that has not been
exploited in quantum mechanics textbooks. For instance, the very
elegant formal treatment discovered by Glauber allows a much
nicer presentation of the gaussian wave packets for a free
particle. The similarities and differences appearing in the
harmonic oscillator and in the free particles are emphasized
providing  a global outlook of quantum mechanics. Another
advantage of the formalism is that one can easily relate the
drift operators with the Galilei group and see that it must be a
\emph{projective} group. The expansion in terms of Hilbert space
bases related with observables of position, momentum and number
are exposed and compared with the expansion in the over-complete
set of coherent states that provide a natural way for introducing
the Husimi distribution. The paper is written with complete
calculations, or with sufficient hints for the calculations, in a
way that it can be directly handed to students as complement of
any textbook.


\begin{thebibliography}{99}
\bibitem{schr} E. Schr{\"o}dinger, Naturwissenschaften \textbf{14}, 664 (1926).

\bibitem{glau}R. J. Glauber ``Coherent and Incoherent States of the Radiation
Rield'' Phys. Rev. \textbf{131}, 2766 (1966).

\bibitem{dlt}
A. C. de la Torre and D. Goyeneche ``Quantum mechanics in
finite-dimensional Hilbert space'' Am. J. Phys., \textbf{71},
49-54, (2003).

\bibitem{dlt3}A. C. de la Torre, P. Catuogno, S. Ferrando, ``Uncertainty
and nonseparability'', Found. of Phys. Lett. {\bf 2}, 235-244,
(1991).

\bibitem{dlt1}
A. C. de la Torre ``On Randomness in Quantum Mechanics'' Eur. J.
Phys. \textbf{29}, 567-575, (2008).

\bibitem{ball}L. E. Ballentine, \emph{``Quantum Mechanics. A Modern
Development''}, World Scientific, Singapore, (1998).

\bibitem{lee}H. W. Lee ``Theory and application of the quantum
phase-space distribution functions'' Phys. Rep.  {\bf 259},
147-211,(1995).
\end{thebibliography}
\end{document}